# WHY DO ONLY SOME RADIO GALAXIES ACQUIRE GIANT SIZES?

Elizabeth López Vázquez[1]  and  Heinz Andernach[2]

## RESUMEN

Se presenta un estudio de la morfología de la radioemisión de radiogalaxias *gigantes* (GRGs), una especie muy rara de galaxias activas, para encontrar nuevos indicios de las razones de su gran tamaño. En base a imágenes de dos rastreos del cielo a 1.4 GHz para cuantificar su radiomorfología, medimos la geometría (simetría, ángulo de distorsión, etc.) y la razón entre flujos de cada lóbulo, en 58 de tales objetos. Un análisis preliminar sugiere que la simetría de las radiofuentes no difiere entre galaxias y cuásares, además no hay una tendencia entre tamaño lineal y corrimiento al rojo, y sólo una tendencia marginal para mayor simetría con mayor tamaño. Integrando datos disponibles de otros objetos se espera obtener nuevos indicios sobre las razones de su gran tamaño. Además, en estos rastreos inspeccionamos alrededor de 1059 galaxias y cuásares, en búsqueda de posibles nuevos ejemplos de radiocuásares gigantes y se presenta una muestra preliminar de seis de los mismos.

## ABSTRACT

We study the morphology of the radio emission of *giant radio galaxies* (GRGs), a rare type of active galaxies, in order to find new clues for the reasons of their large size. Using radio images from two sky surveys at 1.4 GHz we quantified their radio morphology by measuring the geometry (armlength, bending angle, etc.) and flux symmetry for 58 such objects. Preliminary analysis suggests that radio source symmetry does not differ between galaxies and quasars, that there is no evidence for a decrease in linear size with redshift, and only a marginal trend for increasing symmetry with larger size. A merging with data available for other such objects is expected to yield new clues on the possible reasons for their large size. We also searched these radio surveys around the positions of 1059 galaxies and quasars for further giant radio sources, and present a preliminary sample of six of these.

**Palabras Clave:** astrofísica extragaláctica, galaxias, radiogalaxias, cuásares.

## INTRODUCTION

Giant Radio Galaxies (GRGs) or quasars (GRQs) are those whose radio emission extends over a (projected) largest linear size (LLS) of at least 1 Mpc (or $3.09 \times 10^{22}$ m). They are a small subsample of radio-emitting *active galactic nuclei* (AGN) with a supermassive black hole (SMBH) at their center. Since 2012, when only about 100 GRGs were known, our research group (e.g. Andernach et al., 2012), together with volunteers of the citizen science project *Radio Galaxy Zoo* (http://radio.galaxyzoo.org), more than quadrupled this number. In order to study the possible relation between the optical activity of the galaxy nuclei with the size and shape of their radio emission, Andernach et al. (2015) had used a preliminary sample of 193 GRGs with optical spectra available from the *Sloan Digital Sky Survey* (SDSS, Alam et al., 2015) to conclude that GRGs can have any activity type, but that the activity changes from dominantly quasars at high redshift ($z > 0.4$) to dominantly low-luminosity AGNs at low redshift ($z < 0.4$).

For these GRGs some radio-morphological parameters like armlength ratio (of the distance between central object and the radio sources on each side), bending angle (the orientation difference between the vectors from the central object to the radio source), the fluxes and sizes of each lobe, etc., had been measured by Jiménez Andrade (2015), but had not been included in his analysis. In the meantime, another 58 GRGs with spectra from SDSS had been found, for which radio parameters were measured in the present work. In an attempt to search for further GRGs, we also inspected two samples of altogether 1123 quasars or quasar candidates for the presence of extended radio emission. To convert the largest angular size (LAS) to physical sizes (LLS), and to derive radio luminosities, we use standard cosmological parameters of $H_0 = 70$ km s$^{-1}$ Mpc$^{-1}$, $\Omega_m = 0.3$, and $\Omega_\Lambda = 0.7$.

---

[1] Universidad de Guanajuato, Departamento de Letras Hispánicas, DCSyH; Ex-Convento de Valenciana s/n; Col. Valenciana, C.P. 36240, Guanajuato, GTO, n_imbruglia@hotmail.com

[2] Universidad de Guanajuato, Departamento de Astronomía, DCNE; Cjón. de Jalisco s/n, Col. Valenciana, C.P: 36240, Guanajuato, GTO,  heinz@astro.ugto.mx

## METHODS AND MATERIALS

Radio images at 1.4 GHz from the *NRAO VLA Sky Survey* (NVSS; Condon, Cotton, Greisen et al., 1996) and from the *Faint Images of the Radio Sky* (FIRST; Becker, White & Helfand, 1995), with angular resolutions or *beam widths* of 45" and 5.4", respectively, were displayed side-by-side within the *Aladin* software (Bonnarel, Fernique, Bienaymé et al., 2000). The angular distance from the host galaxy or quasar to the outer extremes of the radio lobes on both sides, as well as the position angle of these vectors, were measured with *Aladin*'s `dist` option, such that the bending (or misalignment) angle between the lobes could be derived as the complement of the difference of these angles. Whenever a compact bright emission (a so-called *hotspot*) was seen at the end of a lobe in the FIRST image, it was used to measure the armlength, unless the lobe or hotspot was resolved out due FIRST's much higher angular resolution. In the latter case the armlength of the lobe was measured in the NVSS image, drawing a vector from the core to the apparent end of the lobe, which we assumed at a brightness of about five times the noise level, but well below the peak of the lobe emission, so as to avoid overestimation of size due to the smearing effect of the telescope beam. In the NVSS image the radio core was often blended with the lobe emission, in which case the vector was drawn from the image center, which, by definition, was chosen at the position of the host galaxy or quasar.

Total flux densities for radio cores and lobes were measured by integration of the radio images over a contiguous area, hand-drawn using the `draw` option in *Aladin*. This area was chosen typically so as to include an area that exceeded the area of significant lobe emission in NVSS by at least half a beam width. Whenever the radio core was blended with lobe emission in NVSS, the integration area was divided between the lobes by a line through the radio core perpendicular to the source major axis, and from each NVSS lobe flux half of the flux of the radio core, taken from FIRST, was subtracted. Whenever the radio core was well separated from the lobes in either NVSS or FIRST, its flux was measured in both, such that a difference between them may indicate the presence of either a weak jet, or variability of the core itself (since NVSS and FIRST were observed at different epochs). Often, an extended lobe was partially or completely resolved out in FIRST in which case we only integrated the flux over the area with emission noticeably above the noise level. If no such regions were seen in FIRST, we assigned zero flux to that lobe.

An example of our use of *Aladin* is shown in Fig. 1. The NVSS image of GRG J0107+0246 at redshift z=0.196 and with LAS=7.6', corresponding to LLS=1.4 Mpc, is shown in grey scale in the left panel, while the right panel shows the same sky area in the FIRST survey. Both images are centered on the host galaxy, which features a faint radio core in the FIRST image, but it is blended with lobe emission (or possibly faint jets) in the NVSS image. Since the northern lobe is resolved out in the FIRST image we measure its length (of 3.59' from the core to the end of the lobe) in NVSS. The southern lobe has a compact hotspot in the FIRST image, which we consider the end of the lobe. To measure the integrated flux, e.g. of the south lobe, we drew by hand the red curve around its extent in NVSS, which is automatically reproduced in the FIRST image. The flux in Jansky (1 Jy = $10^{-26}$ W m$^{-2}$ Hz$^{-1}$) is then the product of the value of "**Sum**" (as listed in each panel of Fig. 1), times the ratio of pixel-to-beam area (0.098 here).

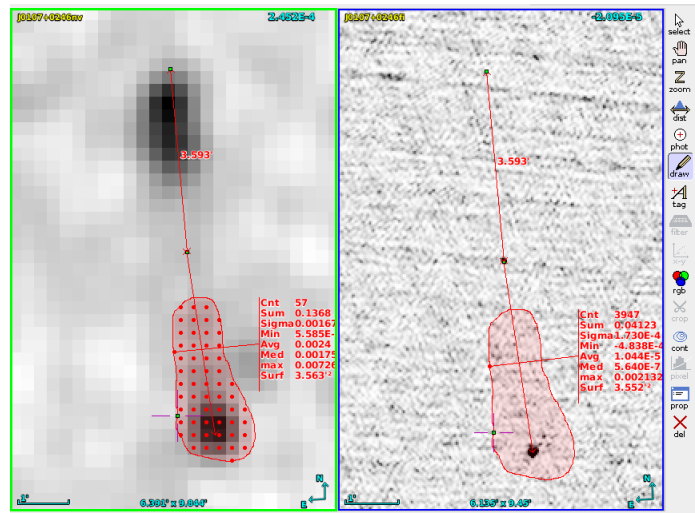

*Fig. 1. Example of using Aladin to determine parameters for giant radio sources (see text for details)*

All of the above parameters were logged in a table, such that the bending angle could be derived from the individual armlength vectors, and the flux ratios between the lobes and the flux ratio between FIRST and NVSS for each lobe. The lower the latter ratio, the more diffuse the emission is, with a ratio of zero corresponding to "genuinely diffuse". In addition, we pursued two exploratory attempts to find more GRGs. The first sample was drawn from table 9 of Kimball et al. (2011) which lists 719 quasar candidates, selected by these authors to have a radio core that coincides with an SDSS object fainter than $i$=19.1 mag, no available redshift in SDSS DR6, and two FIRST sources within 2' from the core, with a bending angle of at most 60°. Since only very few of these objects had measured redshifts, which are required to obtain linear sizes for their radio structures, we culled photometric redshifts ($z_{phot}$) from various sources like Alam et al. (2015), Brescia, Cavuoti, Longo, & DeStefano (2014), DiPompeo, Bovy, Myers, & Lang (2015), and Bilicki, Peacock, Jarrett, et al. (2016). We obtained values of $z_{phot}$ from at least one of these references for 643 (~89%) objects. However, while Kimball et al. (2009) claimed that the 719 objects are ``excellent candidates for optical quasars'', only 253 (or 35%) were found by us among quasar candidates in DiPompeo et al. (2015).

The second sample was drawn from a list of revised redshifts published by Yuan, Strauss & Zakamska (2016) for 482 quasars, which these authors found to be erroneous in SDSS, implying that many of them actually lie at larger distances than previously thought. To investigate whether any of these qualified as a GRG, we searched NVSS images for diffuse emission, and FIRST images to determine the positions of compact sources in the lobes suspected in NVSS. The latter positions were then inspected for the presence of optical objects in SDSS DR12 (http://skyserver.sdss.org/dr12/en/tools/search/radial.aspx), and if the latter was the case, were discarded as GRGs.

**RESULTS**

From NVSS and FIRST images of the 58 new GRGs with SDSS spectra, we recorded (where possible) the armlength of each lobe, its position angle, the maximum angular width of each lobe, their integrated fluxes and the integration areas used (in order to determine their average surface brightness), as well as the flux of the radio core, when present. Each lobe was classified for its so-called FR type, being I if the peak brightness occurred within half the lobe length as measured from the core, and II otherwise, regardless of the core flux (Fanaroff & Riley 1974).

In the left panel of Fig. 2 we show the armlength ratio (longer-to-shorter lobe) as function of LLS (in Mpc) of the 58 radio galaxies (33 Gs, red) and quasars (25 Qs, blue), while in the right panel we show the bending angle between the two lobes as function of LLS. Clearly, Gs and Qs do not segregate in these plots, implying a similar amount of symmetry for both. Only the left panel shows a trend for larger sources being more symmetric in armlength ratio.

The median redshift of Gs and Qs is 0.45 and 0.91, respectively. The distribution of LLS for Gs and Qs, with medians of 1.23 and 1.10 Mpc, respectively, is statistically indistinguishable, as is that of the bending angles (with a median of 5° for both), and that of the FR class distribution (88% for Gs and 84% FRII for Qs). Only 3 of the 58 sources (5%) were found to be of type FRI, and another five cases with different FR types between one lobe and the other (so-called hybrid sources, or HyMORs). Of these latter, four had the *longer* lobe of type FRI. Curiously, there is no trend of LLS with redshift z, despite the fact that at higher z the Universe was denser and should have impeded the growth of radio lobes more than at lower z. It will be interesting to analyze our data with respect to the lobes' surface brightness, since modern cosmology predicts a *surface brightness dimming* proportional to $(1+z)^4$. For the higher-redshift sources (7 of our 58 objects have a z>1) this effect should be noticeable.

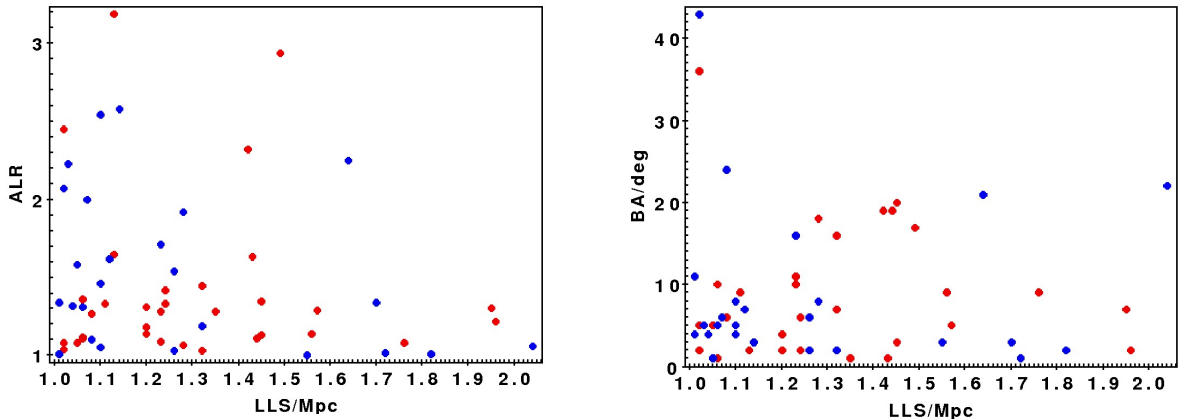

*Fig. 2. Geometrical parameters for the 58 new giant radio galaxies (see text for details).*

Of the 719 objects in table 9 of Kimball et al. (2009), 142 were already listed as extended radio sources in a compilation maintained by one of us (HA). The remaining 577 quasar candidates were searched by us for extended radio structures emanating from these objects in NVSS images of 15' x 15', and, when found, their total angular size was logged. Only those 23 objects with LAS > 3', and with a value for $z_{phot}$ available and suggesting an LLS > 1 Mpc, were inspected on FIRST images, while objects with too asymmetric radio structures, as well as those with optical objects coinciding with the suspected radio lobes, were discarded. As a result, five new giant radio galaxies were found, two of them are shown in Fig. 3. All images are oriented with north up and east to the left, and NVSS contour plots mark the optical host positions as a central cross. The left two panels show the NVSS and FIRST images of 2' x 6' of GRG J1341+0625, hosted by SDSS J134122.39+062504.9 with $z_{phot}$=0.45, LAS=3.78', thus LLS=1.31 Mpc. The right two panels show NVSS and FIRST images of 2.5' x 7' of GRG J1357+6009, hosted by the galaxy SDSS J135732.72+600929.5 with $z_{phot}$=0.35 and LAS=5.35', thus LLS=1.59 Mpc. Both sources are core-dominated with neither of the outer hotspots having an optical or infrared counterpart, and their armlength ratios of 1.09 and 1.8, and bending angles of 10.9° and 7.4°, respectively, are rather typical for GRGs (see e.g. Fig. 2). Three other examples of giants were found: the galaxy SDSS J235804.24-110241.6 with $z_{phot}$=0.49 and LLS=1.37 Mpc, as well as two distant quasars, namely SDSS J154221.45+633741.3, with $z_{phot}$=0.84 and LLS=1.48 Mpc, and SDSS J123824.94+422836.0 with $z_{phot}$=2.86 and LLS=1.45 Mpc. Several dozen more candidates remain to be inspected.

Inspection of NVSS and FIRST images of the 482 quasars from Yuan et al. (2016) yielded over 100 preliminary candidates. Again, the positions of potential outer lobes or hotspots were checked in SDSS DR12 and were discarded when found to coincide with an optical object within ~2". Only a single GRG candidate was found, hosted by SDSS J005504.42+273036.7 with $z_{spec}$=0.576, LAS=3.6', thus LLS = 1.42 Mpc. Unfortunately this object is not covered by the FIRST survey, and is too faint to be detected in the recent TGSS-ADR1 survey at 150 MHz (Intema et al., 2016). The GRG nature of this object thus needs further confirmation from deeper radio imaging.

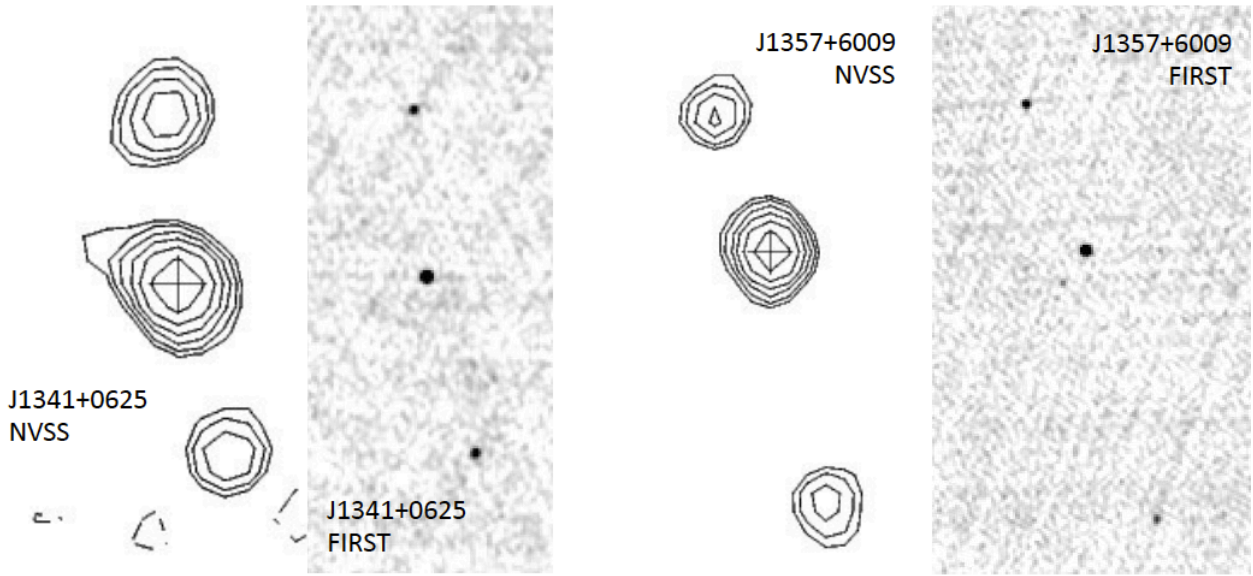

*Fig. 3. Examples of newly discovered giant radio galaxies from Kimball et al. (2009), see text for details.*

## CONCLUSIONS

We measured radio morphological parameters like armlength ratio, bending angle, flux density ratio between longer and shorter lobe, radio luminosity, radio core flux fraction and diffuseness of the lobes, for a new sample of 58 giant radio galaxies (33 galaxies and 25 quasars) which also have an optical spectrum in the SDSS database. We did not find significant differences in radio morphology or size between galaxies and quasars. However, the fact that quasars have a median redshift twice that of the galaxies, implies there is no trend for the size to decrease with redshift.

NVSS and FIRST radio images for 577 quasar candidates from Kimball et al. (2009) and 482 quasars with revised redshifts from Yuan et al. (2016) were searched for extended sources. In the 5 weeks available we could only follow up the most promising cases, finding five convincing GRGs larger than 1 Mpc from the former reference, and one from the latter. The results of this project will be merged with an existing compilation of GRGs by one of us (HA) and will contribute to find the reasons why only a small fraction of radio galaxies reach these enormous sizes.